\documentclass[a4paper]{article}

\usepackage{bm}
\usepackage{array}
\usepackage{multirow}
\usepackage{amsmath}
\usepackage{cite}
\usepackage{graphicx}
\usepackage{balance}
\usepackage{INTERSPEECH2020}

\ninept
\title{Full Attention Bidirectional Deep Learning Structure for Single Channel Speech Enhancement}
\name{Yu-Zi Yan$^1$, Wei-Qiang Zhang$^1$, Michael T. Johnson$^2$
\thanks{The corresponding author is Wei-Qiang Zhang.}
\thanks{This work was supported by the National Natural Science Foundation of China under Grant No. U1836219 and the National Key R\&D Program of China.}}
\address{$^1$Beijing National Research Center for Information Science and Technology \\
  Department of Electronic Engineering, Tsinghua University, Beijing 100084, China \\
  $^2$Electrical and Computer Engineering, College of Engineering \\
  University of Kentucky, Lexington, KY 40506-0046, USA
  }
\email{yanyz17thu@gmail.com, wqzhang@tsinghua.edu.cn, mike.johnson@uky.edu}

\begin{document}

\maketitle
\begin{abstract}
  As the cornerstone of other important technologies, such as speech recognition and speech synthesis,
  speech enhancement is a critical area in audio signal processing.
  In this paper, a new deep learning structure for speech enhancement is demonstrated.
  The model introduces a ``full" attention mechanism to a bidirectional sequence-to-sequence method to make use of latent information after each focal frame.
  This is an extension of the previous attention-based RNN method.
  The proposed bidirectional attention-based architecture achieves better performance in terms of speech quality (PESQ), compared with OM-LSA, CNN-LSTM, T-GSA and the unidirectional attention-based LSTM baseline.

\end{abstract}

\section{Introduction}
Neural networks have seen great success in the field of speech enhancement and noise suppression in the last decade \cite{Narayanan2013Ideal, geiger2014investigating, Erdogan2015Phase, wang2014training, pascual2017segan, rethage2018wavenet, soni2018time, ephraim1985speech}.
This approach now outperforms most traditional model-based statistical approaches,
such as spectral subtraction \cite{boll1979suppression}, the minimum mean-square error log-spectral method \cite{ephraim1985speech}, Wiener filtering \cite{scalart1996speech} or OM-LSA \cite{tran2010speech}.
Recurrent neural network (RNN), has been widely used \cite{chen2017long, weninger2015speech, sun2017multiple}.

Various RNN-based structures such as gated recurrent unit (GRU) networks, long short term memory (LSTM) networks, have been explored due to their powerful sequence learning ability, as well as combinations and variations \cite{valin2018hybrid}, which show superiority in the field of audio signal processing.
The introduction of an attention mechanism further enhances the potential of these seq-to-seq methods \cite{li2019multi}.
The proposed approach makes use of Mel frequency features \cite{wu2000word} to extract features to represent the sound sequence and introduces  a new bidirectional attention-based structure in speech enhancement.

Our work is inspired by the recent success of unidirectional attention-RNN models in various seq-to-seq learning tasks \cite{hao2019attention}, which allow the clean speech to be perceived with high attention while the noise and background have low attention.
We extend this idea to make full use of the information around the focal frame as indicated by a high level of attention, and refer to this as a \emph{full} as opposed to partial or half attention method typical of previous approaches.
The new architecture uses a bidirectional connection structure combined with this full attention mechanism.

The rest of the paper is organized as follows. In Section \ref{sec_method}, our method is described in detail.
The experiments and results are presented in Section \ref{sec_exp}.
Finally we draw conclusions in Section \ref{sec_conclusion}.

\section{Method} \label{sec_method}

\begin{figure*}[htb]
  \centering
  \includegraphics[height=2.3in]{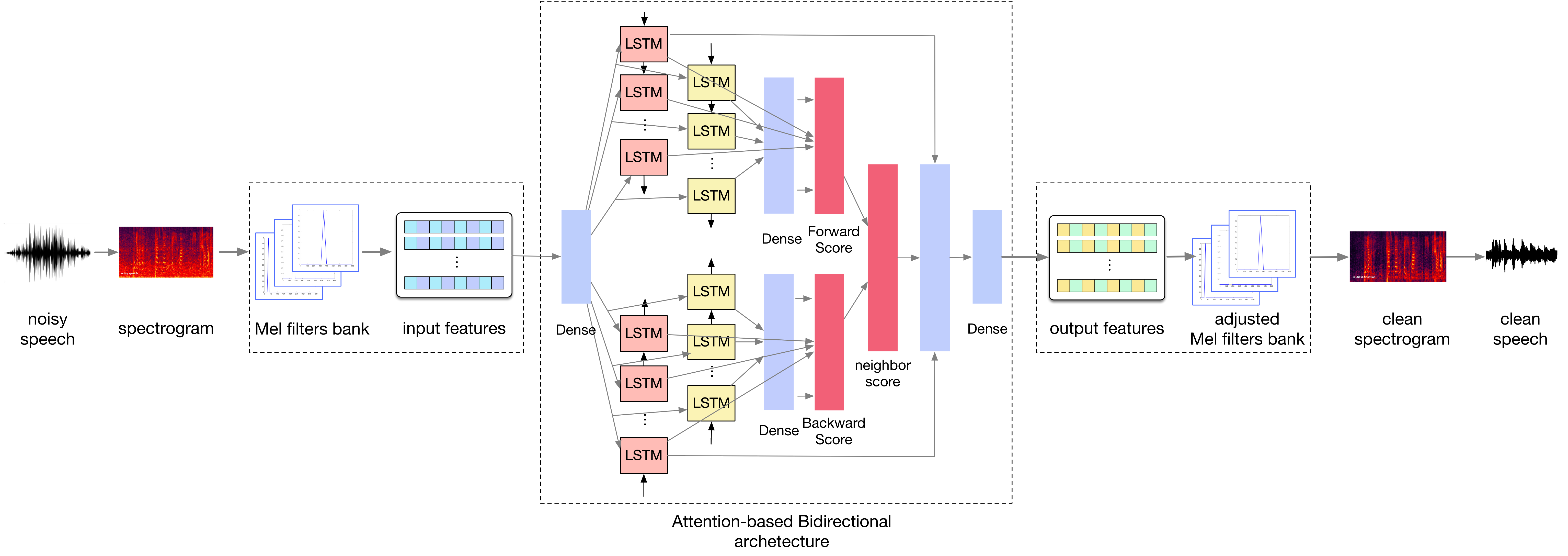}
  \caption{System Overview.} \label{fig1}
\end{figure*}

As illustrated in Figure \ref{fig1}, the process is divided into three major parts: feature extracting, deep learning and filtering.
The input audio signal is speech with noise and the output audio signal is expected to be clean.
Feature extraction calculates Mel filter bank (FBank) features,
while the neural network implements regression and generates clean estimates of the FBank \cite{wu2000word} for filtering.

\subsection{Feature Extraction}

Frequency features are extracted from the short-time Fourier transform (STFT) of input audio signal.
In contrast to typical methods that use all sampling points in the spectrum as features \cite{hao2019attention},
we choose to extract the features in FBank to reduce the number of the weight to estimate in the system.
We use the input and output feature vectors to conduct spectrum suppression as described in Section \ref{subsec_mel}.

\subsection{Attention-based Bidirectional Architecture}

\begin{figure}[tbh]
  \centering
  \includegraphics[width=2.8in]{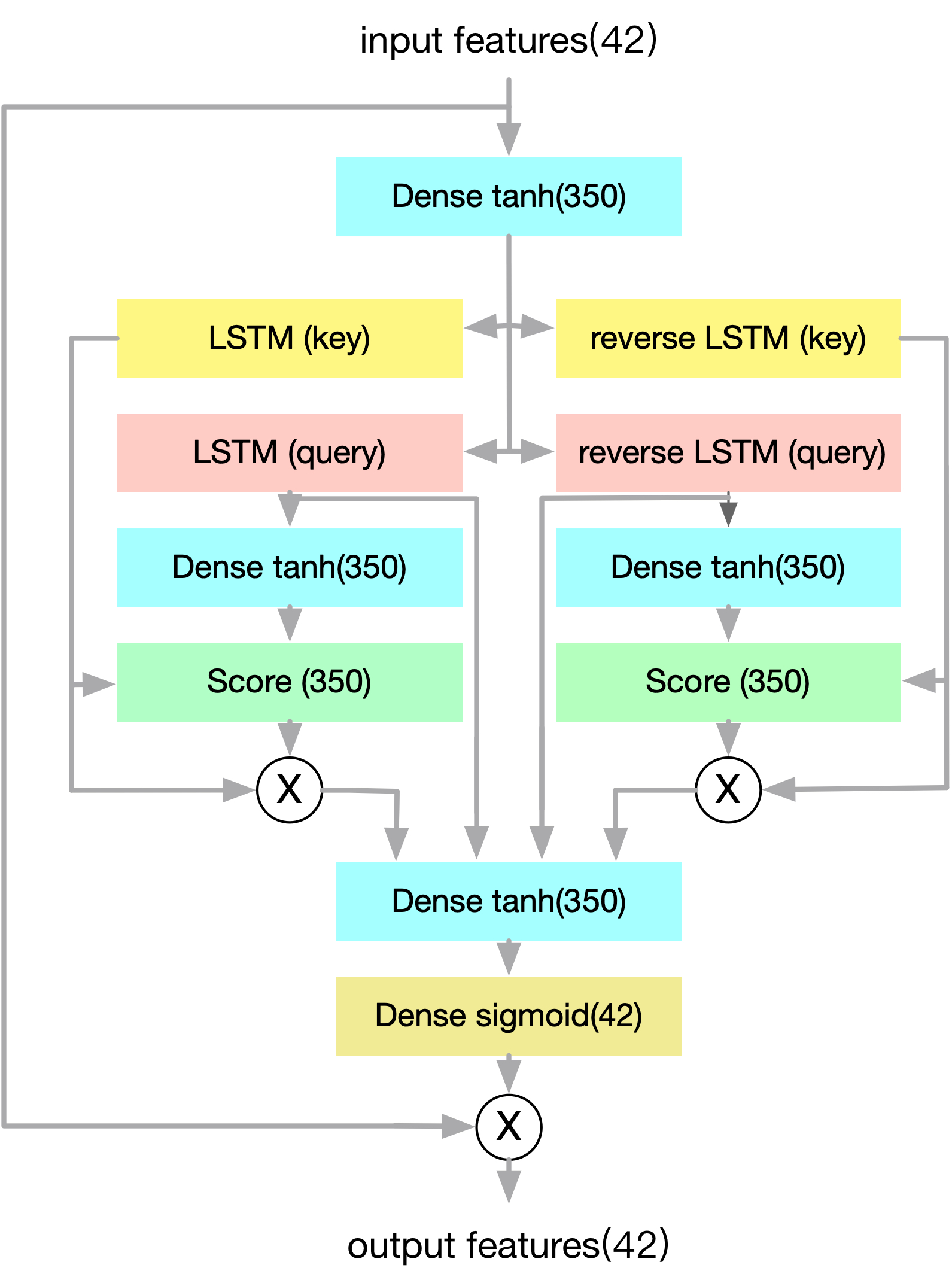}
  \caption{Deep Learning Structure: The left and right sides respectively represent the calculation process of the forward and reverse attention parameters.} \label{fig2}
\end{figure}

The deep learning architecture is shown in Figure \ref{fig2}.
The input is $\bm{X}=[\bm{x}_1, \bm{x}_2, ... \bm{x}_T]$, where $\bm{x}_t$ is the vector that represents the amplitude at specific frequency points of noisy speech at frame $t$. $T$ represents the total number of frames.

A dense layer serves as the encoder to give a high-level representation from the input features.
\begin{equation}
\bm{\hat{x}}_t = \tanh{(\bm{W} \bm{x}_t+\bm{b})}
\end{equation}
The number of output features requires fine adjustment to maintain balance between representation and generalization.
$\bm{\hat{x}}_t$ is used in both the forward LSTM cell and the backward LSTM cell to generate the key and the query vectors.
\begin{align}
\bm{h}_{t,f}^K &= {\rm{LSTM_1}}({\rm{forward}}(\bm{\hat{x}}_t)) \\
\bm{h}_{t,b}^K &= {\rm{LSTM_2}}({\rm{backward}}(\bm{\hat{x}}_t)) \\
\bm{h}_{t,f}^Q &= {\rm{LSTM_3}}({\rm{forward}}(\bm{\hat{x}}_t)) \\
\bm{h}_{t,b}^Q &= {\rm{LSTM_4}}({\rm{backward}}(\bm{\hat{x}}_t))
\end{align}
The key and query vectors are shown as $\bm{h}_{t,f}^K$, $\bm{h}_{t,b}^K$, $\bm{h}_{t,f}^Q$ and $\bm{h}_{t,f}^Q$,
where the backward or forward direction is denoted by subscript $b$ or $f$, respectively.
As shown above, backward LSTM cell can be regarded as a normal LSTM cell with the reverse sequence input.

This approach will make it possible to use latent information in the attention mechanism.
A dense layer is also added before generating the final query, but this step is omitted here for the simplicity of explanation.
In forward and backward processing, the attention mechanism generates two attention weight vectors.

\begin{align}
\bm{\alpha}_t &= [\alpha_{t,t-\omega}, \alpha_{t,t-\omega+1}, ... ,\alpha_{t,t}] \\
\bm{\gamma}_t &= [\gamma_{t,t+\xi}, \gamma_{t,t+\xi-1}, ... ,\gamma_{t,t}]
\end{align}
Following the correlation calculation in \cite{hao2019attention} and \cite{luong2015effective}, normalized attention weight $\alpha$ and $\gamma$ can be learned as:
\begin{align}
\alpha_{t,k} &= \frac{\exp{(\bm{h}_{k,f}^K \bm{W} \bm{h}_{t,f}^Q)}}{\sum_{k=t-\omega}^t\exp{(\bm{h}_{k,f}^K \bm{W} \bm{h}_{t,f}^Q})} \\
\gamma_{t,k} &= \frac{\exp{(\bm{h}_{k,b}^K \bm{W} \bm{h}_{t,b}^Q)}}{\sum_{k=t}^{t+\xi}\exp{(\bm{h}_{k,b}^K \bm{W} \bm{h}_{t,b}^Q)}}
\end{align}

The attention weight vectors indicate the degree of relevance of neighboring frames to the time frame $t$, which we refer to as a \emph{focal frame}.
If the utterance is too long, the attention weights of distant frames may be nearly zero.
Based on human pronunciation patterns, it is reasonable to assume that the length of backward effective attention sequences and the length of forward effective attention sequences may be different.
Based on thie, $\omega$ and $\xi$ are implemented as separate constants.

If $t-\omega<0$, we can set $t-\omega = 0$. If $t+\xi>N$, we can set $t+\xi=N$, where $N$ is the total number of frames in the input sequence.
The context vector $\bm{c}_t$ containing the information of key vectors $\bm{h}_{k,f}^K$ and $\bm{h}_{k,b}^K$ can be computed.
\begin{align}
\bm{c}_{t,f} &= \sum_{k=t-\omega}^t(\alpha_{t,k}\bm{h}_{k,f}^K) \\
\bm{c}_{t,b} &= \sum_{k=t}^{t+\xi}(\gamma_{t,k}\bm{h}_{k,b}^K) \\
\bm{c}_t &= [\bm{c}_{t,f};\bm{c}_{t,b}]
\end{align}
where $[\cdot;\cdot]$ denotes the concatenation of two vectors.
Another dense layer severs as a decoder to make use of the context vector $\bm{c}_t$, query vectors $\bm{h}_t^Q = [\bm{h}_{t,f}^Q; \bm{h}_{t,b}^Q]$ and the input feature $\bm{x}_t$.
Similar to \cite{hao2019attention}, the model generate an enhancement vector $\bm{e}_t$ and the final gain $\bm{g}_t$ as:
\begin{align}
\bm{e}_t &=  \tanh{(\bm{W}_e[\bm{c}_t; \bm{h}_t^Q]+\bm{b}_e)} \\
\bm{g}_t &= {\rm{sigmoid}}(\bm{W}_m \bm{e}_t+\bm{b}_m) \\
\bm{y}_t &= \bm{x}_t \odot \bm{g}_t
\end{align}

\subsection{Mel Filter Bank Generator} \label{subsec_mel}

The model generates a FBank feature vector for every frame to use for noise filtering.
The gain $\bm{g}_t$ of each triangular filter at the peak response is calculated from the input vector $\bm{x}_t$ and the output vector $\bm{y}_t$.
The amplitude of other points of the filter bank is then adjusted proportionally according to the gain at the peak.

\section{Results} \label{sec_exp}
\subsection{Datasets}
We tested the quality of the speech enhancement on two free speech databases, THCHS-30 \cite{wang2015thchs} and QUT-NOISE-TIMIT \cite{dean2010qut}.

THCHS-30 is a free Chinese speech database, which involves more than 30 hours of clean speech signals recorded by a single microphone in silence and with white noise, car noise and cafeteria noise.
We extend the noise conditions further by creating a mix of car noise and cafeteria noise, implemented along with the white noise condition. Signal-to-noise ratio (SNR) ranges from $-10$ dB to 10 dB.

QUT-NOISE-TIMIT is synthesized by mixing 5 different background noise sources with TIMIT \cite{garofolo1993darpa}.
For the training set, $-5$ and 5 dB SNR data were used, but the evaluation set contains SNR ranging from $-10$ dB to 15 dB.
The total length of train and test data corresponds to 25 hours and 12 hours, respectively.

\subsection{Experiment Setup}
The sampling rate of the audio files is 16 kHz, with 512 point Hanning-windowed frames and 128 points overlap.

In both datasets, we randomly divide the data into training set and test set in a 4:1 ratio.
The result is verified using 5-fold cross-validation.
The input and the output are both 42-dimensional FBank vectors.
The batch size is set to 96,
the number of LSTM cells is set to 350,
and the value of $\omega$ and $\xi$ are set to 15 and 5, respectively.
The loss function is mean square error (MSE), and dropout regularization is used.
The setting of the learning rate is dynamically adjusted according to the SNR as follows:
$$
  \rm LearningRate=\left\{
  \begin{array}{rcl}
  1\times 10^{-6} & & {\rm{SNR} \geq 10dB}\\
  1\times 10^{-5} & & {\rm 5dB \leq \rm{SNR} < 10dB}\\
  5\times 10^{-5} & & {\rm 0dB \leq \rm{SNR} < 5dB}\\
  1\times 10^{-4} & & {\rm -5dB \leq \rm{SNR} < 0dB} \\
  5\times 10^{-4} & & {\rm{SNR} < -5dB}
  \end{array} \right.
$$

In the experiments on the THCHS-30 dataset,
we compared our approach with OM-LSA, LSTM-RNN and LSTM with attention(LSTM-Att) \cite{hao2019attention}.
It is worth mentioning that the LSTM-Att method is unidirectional, which can make the comparision more convincing.

In the experiments on the QUT-NOISE-TIMIT dataset,
we choose CNN-LSTM, O-T, T-AB, T-GSA, C-T-GSA to be the baseline \cite{kim2019transformer}, in order to compare our model with other end-to-end sequence structures.
The perceptual evaluation of speech quality (PESQ) \cite{rix2001perceptual} is used as the evaluation criteria.

\subsection{Main Results}
Results for THCHS-30 are shown in Table \ref{tab1}.
The new approach outperforms the three baselines in most instances, especially when the noise pattern is irregular and SNR is relatively low.
The bidirectional model performs better than the unidirectional model or the model without an attention mechanism.
This sheds light on the effectiveness of using additional context when implementing the attention mechanism.

\begin{table*}[tbh]
  \centering
  \caption{PESQ results on THCHS-30: Test set consists of 5 SNR ranges:-10, -5, 0, 5, 10dB. Bold indicates best results.} \label{tab1}
	\begin{tabular}{l r r r r r r r r r r}
    \toprule
    \multirow{2}{*}{Method}&\multicolumn{2}{c}{Raw}&\multicolumn{2}{c}{OM-LSA}&\multicolumn{2}{c}{LSTM-RNN}&\multicolumn{2}{c}{LSTM-Att}&\multicolumn{2}{c}{\bf Bi-Att}\cr
    &cafe\&car&white  &cafe\&car&white  &cafe\&car&white  &cafe\&car&white &cafe\&car&white\cr
    \midrule
    $-10$ dB&0.6424&0.0310  &0.8337&0.8258  &1.3532&1.3750  &1.3786 &1.3756  & $\bm{1.5615}$& $\bm{1.5044}$\cr
    $-5$ dB &1.1098&0.4169  &1.2895&1.4132  &1.7820&1.6757  &1.8277 &1.7512  & $\bm{1.8317}$& $\bm{1.7986}$\cr
    0 dB  &1.5216&0.9977  &2.1160&1.9523  &2.2624&2.2322  &2.4584 &2.2324 & $\bm{2.5195}$& $\bm{2.2745}$\cr
    5 dB  &2.1597&1.6791  &2.2753&$\bm{2.4582}$  &2.4723&2.4023  &2.6439 &2.4330  & $\bm{2.6675}$&2.4468\cr
    10 dB &2.6210&2.3163  &2.6620&$\bm{2.8326}$  &2.7027&2.4732  &2.8659 &2.4942  & $\bm{2.9185}$&2.5165\cr
    \bottomrule
  \end{tabular}
\end{table*}

Results for QUT-NOISE-TIMIT are shown in Table \ref{tab2}.
The bidirectional attention model outperforms the selected baseline models in most SNR situation.
The area where the improvement is most significant is in the SNR range from $-5$ to 5 dB.
Our hypothesis is that the FBank structure can better suppress the full frequency band under high noise conditions,
but that in the case of low noise energy, insufficiently fine point-to-point suppression may reduce the quality of the entire speech signal.

\begin{table}[tbh]
  \centering
  \caption{PESQ results on QUT-NOISE-TIMIT: Test set consists of 6 SNR ranges: $-10$, $-5$, 0, 5, 10, 15 dB, according to result in \cite{kim2019transformer}.} \label{tab2}
  \footnotesize
  \begin{tabular}{l c c c c c c}
    \toprule
    SNR (dB) & $-10$ & $-5$ & 0 & 5 & 10 & 15 \\
    \midrule
    Raw & 1.07 & 1.08 & 1.13 & 1.26 & 1.44  &1.72    \\
    CNNLSTM  & 1.43 & 1.65 & 1.89 & 2.16 & 2.35 & 2.54       \\
    O-T & 1.29 & 1.45 & 1.63 & 1.87  & 2.07 & 2.29    \\
    T-AB & 1.49 & 1.67 & 1.85 & 2.01 & 2.28 & 2.50       \\
    T-GSA & 1.54 & 1.76 & 2.00 & 2.28 & $\bm{2.51}$ & 2.74          \\
    C-T-GSA& 1.43 & 1.64 & 1.88 & 2.17 & 2.40 & 2.67          \\
    Bi-Att & $\bm{1.56}$ & $\bm{1.84}$ & $\bm{2.09}$ & $\bm{2.35}$ & 2.47 & $\bm{2.79}$         \\
    \bottomrule
  \end{tabular}
\end{table}

\begin{figure}[tbh]
  \centering
  \includegraphics[totalheight=2.3in]{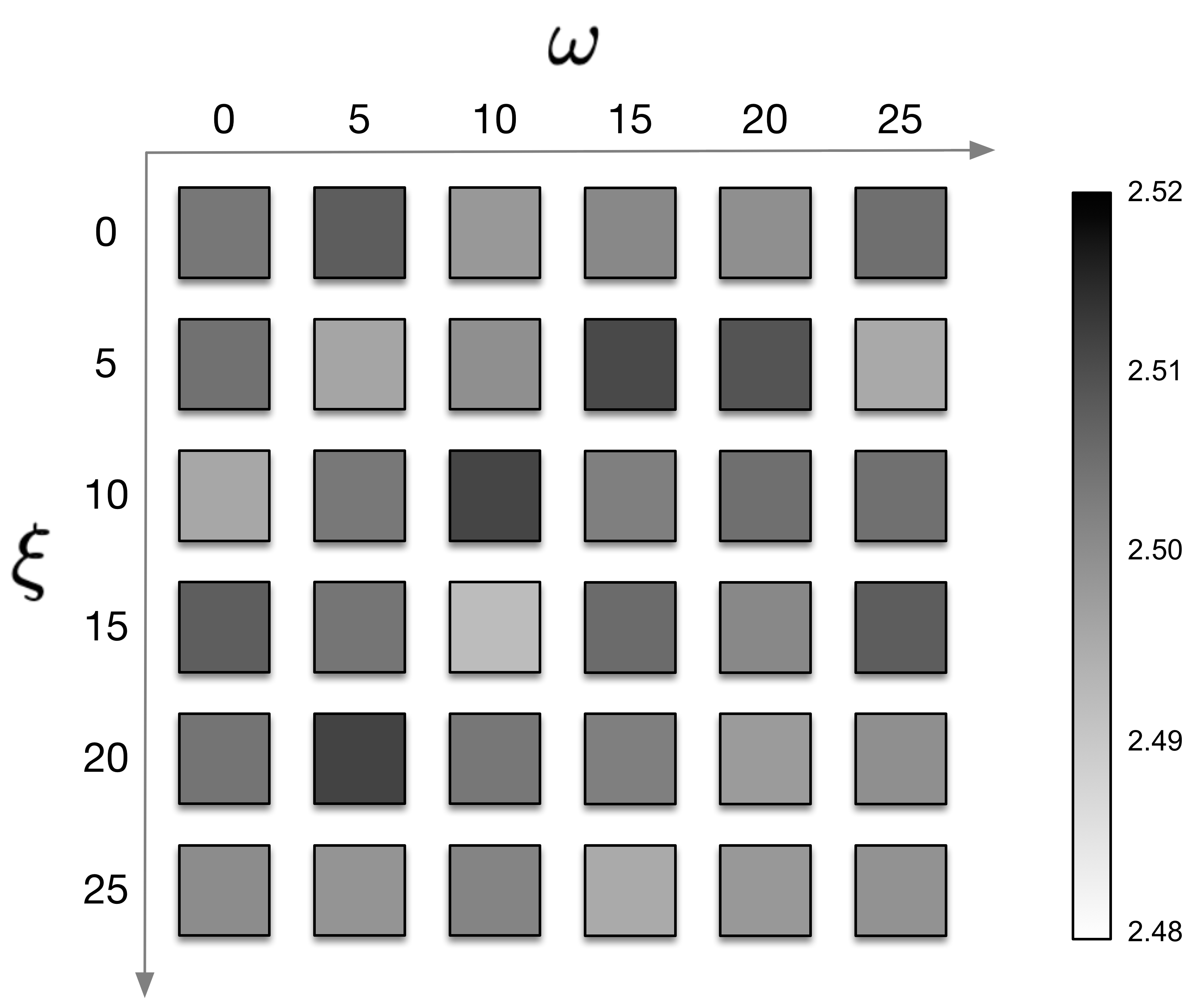}
  \caption{PESQ at different values of $\omega$ and $\xi$ and visualization of results: darker colors represent larger PESQ.
  $\omega$ and $\xi$ represents the length of forward and backward attention sequence respectively.} \label{fig3}
\end{figure}

Based on the asymmetry of speech in the time domain, it is reasonable to assume that the attention may vary in different ways in forward and backward directions.
The results of having separate attention sequence length constants, i.e. $\omega$ and $\xi$, are shown in Figure \ref{fig3}, for the THCHS-30 dataset in mixed car and cafe noise at an SNR of 0 dB.
The resulting PESQ changes slightly as $\omega$ and $\xi$ vary.
We notice that bigger values of $\omega$ and $\xi$ result in no obvious improvement.
The model performs best when the value of $\omega$ and $\xi$ are both in the range of 5 to 20 dB.

\begin{figure}[tbh]
  \centering
  \includegraphics[width=3.1in]{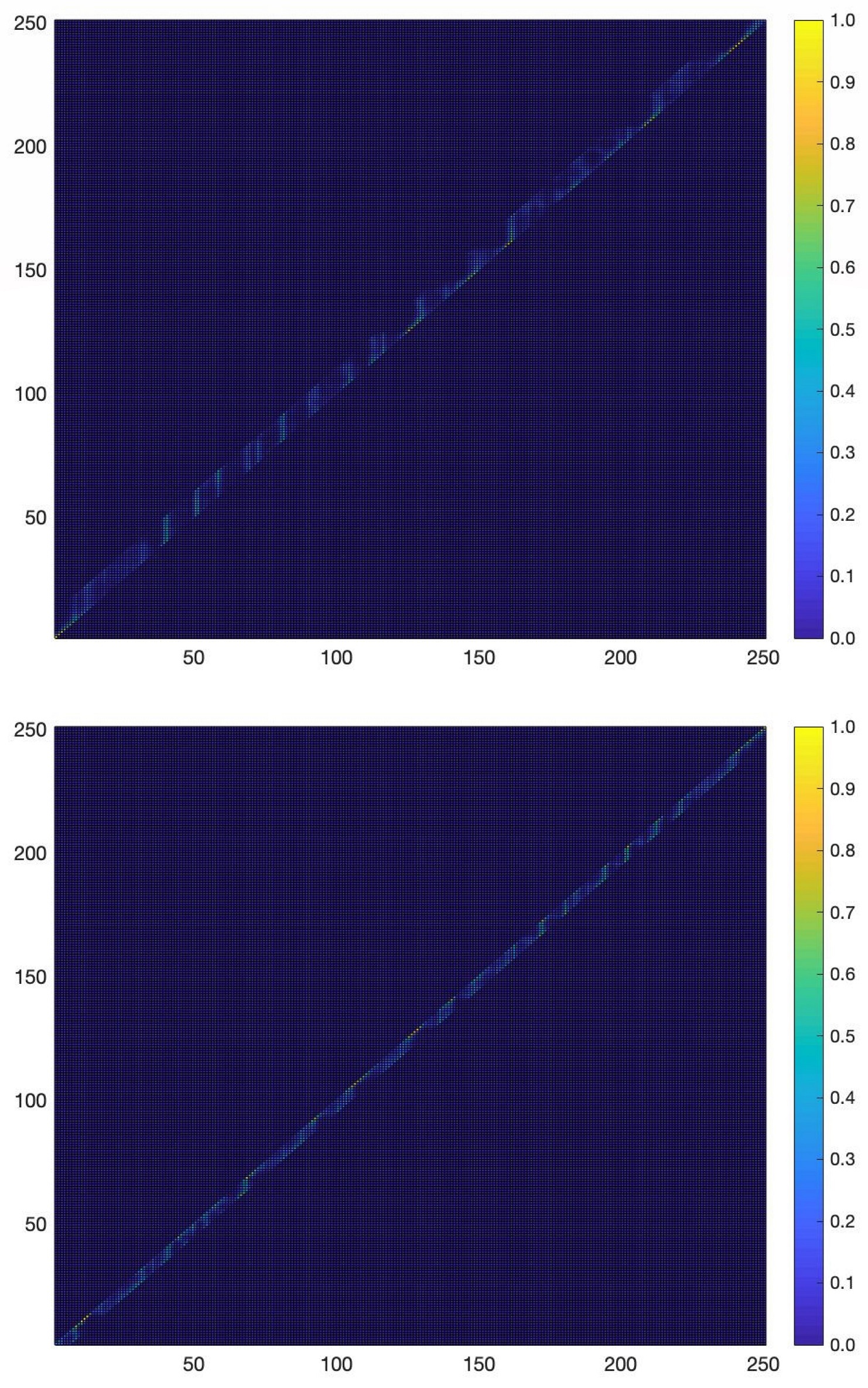}
  \caption{The visualization of bidirectional attention weights, $\omega=15$, $\xi=5$. Top: positive direction. Bottom: negative direction.} \label{fig4}
\end{figure}

\begin{figure}[tbh]
  \centering
  \includegraphics[width=2.8in]{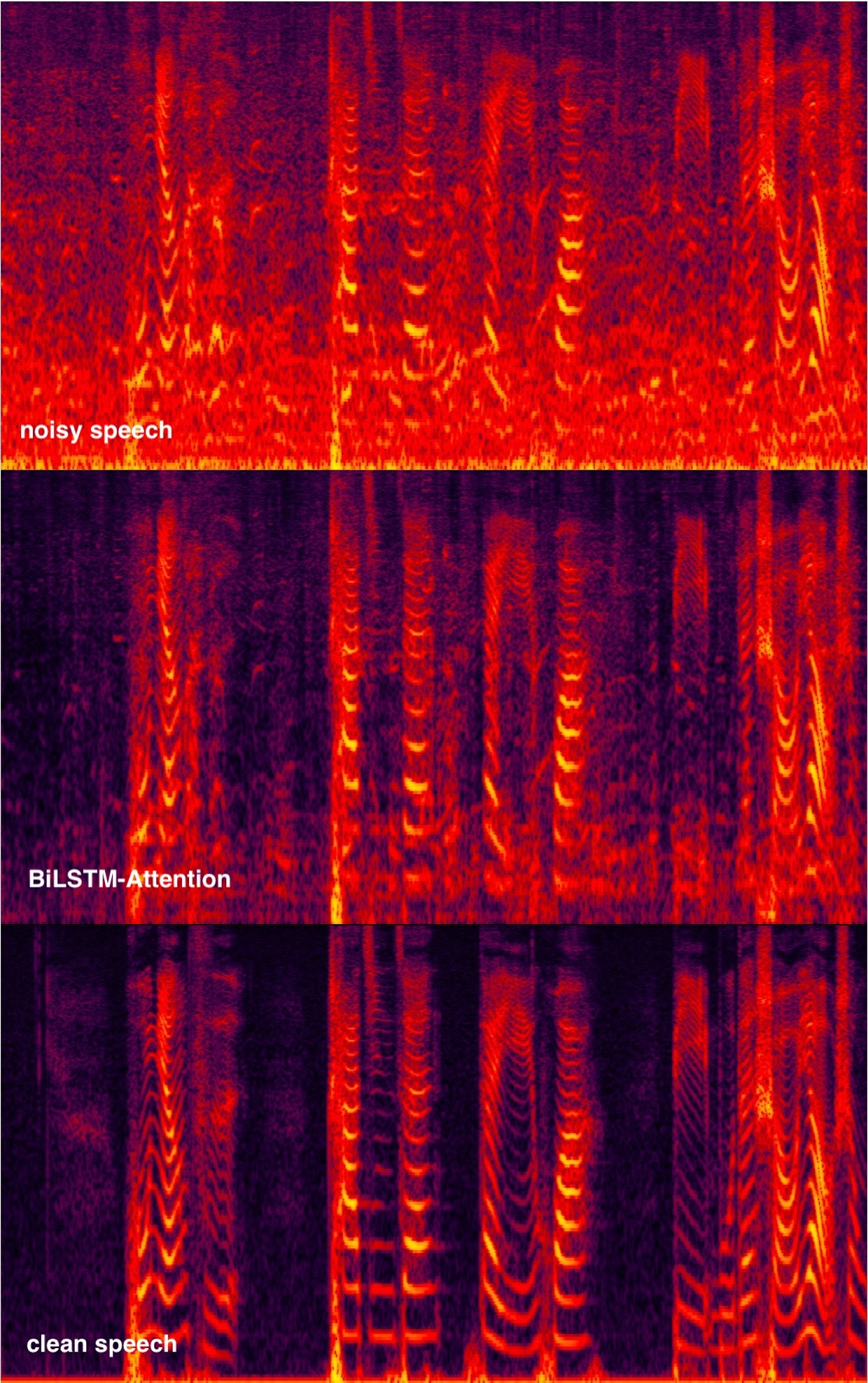}
  \caption{A test example before and after speech enhancement. Top: noisy speech. Middle: speech processed by our model. Bottom: clean speech.} \label{fig5}
\end{figure}

The impact of the attention weight is illustrated in Figure \ref{fig4}.
We set $\omega$ as $15$ and $\xi$ as $5$ on a test example on THCHS-30 (251 frames).
In the left figure, the $x$ and $y$ axis represents $\bm{h}_{t,f}^K$ and $\bm{h}_{t,f}^Q$.
On the right, the $x$ and $y$ axis represents $\bm{h}_{t,b}^K$ and $\bm{h}_{t,b}^Q$.
The point $(x,y)$ represents attention weight.
As seen in the figure, the model gives different weights to the contextual frames.
For illustration, the spectrum of a speech utterance is shown in Figure \ref{fig5}.
The original speech signal is superimposed with the caf\'e and car noise to form noisy speech with an SNR of 0 dB.

\section{Conclusion} \label{sec_conclusion}
This paper demonstrates a new attention-based LSTM architecture for speech enhancement.
The primary innovation is the use of a two-way attention network to take full advantage of latent information.
Mel filter bank features are used to reduce complexity.
Results demonstrate that the resulting perceptual quality of the enhanced speech is higher than comparative baseline methods in most cases.
The attention-based approach also shows better generalization ability to irregular noise conditions.
In addition, asymmetric forward and backward attention weights are implemented and evaluated.
Overall, the proposed approach demonstrates the positive impact of increasing the extent of speech and language context for noise suppression applications.

\balance
\bibliographystyle{IEEEtran}

\bibliography{template}

\end{document}